\documentclass[cits]{PoS}

\usepackage{amsmath}		
\usepackage{amssymb}		
\usepackage{amsthm}		

\usepackage{multirow}		
\usepackage{array}		

\usepackage{mathrsfs}		
\usepackage{bm}			
\usepackage{cancel}		

\usepackage{wrapfig}

\newcommand{\resizedgraphics}[2]{
\resizebox{#1\textwidth}{!}{\includegraphics{#2}}
}

\usepackage{url}			
\usepackage{makeidx}		

\newcommand{\Eq}[1]{Eq.\,\eqref{#1}}
\newcommand{\Fig}[1]{Fig.~\ref{#1}}

\renewcommand{\Re}{\operatorname{\mathfrak{Re}}}
\renewcommand{\Im}{\operatorname{\mathfrak{Im}}}
\newcommand{\Tr}{\operatorname{\textrm{Tr}}}

\newcommand{\abs}[1]{\left|#1\right|}
\newcommand{\ave}[1]{\left\langle#1\right\rangle}
\newcommand{\Ave}[1]{\left\langle\left\langle#1\right\rangle\right\rangle}
\newcommand{\RnB}[1]{\left(#1\right)}
\newcommand{\ClB}[1]{\left\{#1\right\}}

\newcommand{\Nf}{N_{\textrm{f}}}			
\newcommand{\Nsite}{N_{\textrm{site}}}	

\newcolumntype{G}{>{$ \displaystyle \raisebox{15pt}[15pt][10pt]}c<{$}}	
\newcolumntype{C}{>{$ \displaystyle}c<{$}} 							
\newcolumntype{L}{>{$ \displaystyle}l<{$}}  							
\newcolumntype{R}{>{$ \displaystyle}r<{$}}								

\makeatletter
\renewenvironment{thebibliography}[1]
{\section*{\refname\@mkboth{\refname}{\refname}}%
  \list{\@biblabel{\@arabic\c@enumiv}}%
       {\settowidth\labelwidth{\@biblabel{#1}}%
        \leftmargin\labelwidth
        \advance\leftmargin\labelsep
 \setlength\itemsep{1pt}			
 \setlength\baselineskip{11pt}	
        \@openbib@code
        \usecounter{enumiv}%
        \let\p@enumiv\@empty
        \renewcommand\theenumiv{\@arabic\c@enumiv}}%
  \sloppy
  \clubpenalty4000
  \@clubpenalty\clubpenalty
  \widowpenalty4000%
  \sfcode`\.\@m}
 {\def\@noitemerr
   {\@latex@warning{Empty `thebibliography' environment}}%
  \endlist}
\makeatother

\title{Histogram method in finite density QCD with phase quenched simulations}

\ShortTitle{Histogram method in finite density QCD}

\author{\speaker{Y.~Nakagawa}, S.~Ejiri \\
        Graduate School of Science and Technology, Niigata University, Niigata 950-2181, Japan \\
        E-mail: \email{nakagawa@muse.sc.niigata-u.ac.jp}
        }

\author{S.~Aoki, K.~Kanaya, H.~Ohno%
        \footnote{
        Current address: Fakult\"{a}t f\"{u}r Physik, Universit\"{a}t,
        Bielefeld, D-33501 Bielefeld, Germany
        },~
        H.~Saito \\
        Graduate School of Pure and Applied Sciences, University of Tsukuba,
        Tsukuba, Ibaraki 305-8571, Japan
        }

\author{T.~Hatsuda \\
        Department of Physics, The University of Tokyo, Tokyo 113-0033, Japan
        }

\author{T.~Umeda \\
        Graduate School of Education, Hiroshima University, Hiroshima 739-8524, Japan
        }

\author{(WHOT-QCD collaboration)}

\abstract{
We propose a new approach to finite density QCD based on
a histogram method with phase quenched simulations at finite chemical potential.
Integrating numerically the derivatives of the logarithm of the quark determinant
with respect to the chemical potential, we calculate the reweighting factor
and the complex phase of the quark determinant.
The complex phase is handled with a cumulant expansion to avoid the sign problem.
We examine the applicability of this method.
}

\FullConference{The XXIX International Symposium on Lattice Field Theory - Lattice 2011 \\
                July 10-16, 2011 \\
                Squaw Valley, Lake Tahoe, California}

\begin{document}

\section{Introduction}
\vspace{-2mm}

In order to reveal the phase structure of QCD,
which is relevant to the study of the the early universe,
the core of the neutron star and the heavy ion collisions,
it is indispensable to study QCD by first principle lattice simulations.
The lattice simulations, however, have the notorious sign problem
at non-zero quark chemical potential $\mu$.
Although there have been proposed several approaches to study finite density QCD,
a satisfactory method reliable at large quark chemical potential is still lacking.
We propose a new approach to finite density QCD
by means of the histogram method and the reweighting technique
\cite{Ejiri:2007ga,Saito:2011fs}
together with phase quenched simulations,
in which the Monte Carlo ensemble is generated without the complex phase of
the quark determinant.
The complex phase is handled with a cumulant expansion to evade the sign problem.
In this report, we examine the applicability of our method;
in particular, we discuss the overlap problem
and the convergence of the cumulant expansion.

\vspace{-3mm}
\section{Histogram method}
\vspace{-1mm}

In order to calculate thermodynamic quantities such as the pressure,
we need to calculate the expectation values of the plaquette and the quark determinant.
If we are interested in quantities which depend only on them,
the histogram method enables us to evaluate the expectation values
from the probability distribution function of the plaquette and the quark determinant.
Here, we discuss the case of the degenerate $\Nf$ flavor case.
An extension to the non-degenerate case is straightforward.
We label the gauge configurations by the space-time averaged plaquette $P$
and the absolute value of the quark determinant,
$F(\mu) = \Nf \ln \abs{\det M(\mu)/\det M(0)}$.
Decomposing the quark determinant as
$\RnB{\det M(\mu)}^{\Nf} = e^{i\theta(\mu)}\abs{\det M(\mu)}^{\Nf}$,
the partition function normalized at zero chemical potential can be written as
\begin{equation}\label{eq:part_func}
\frac{Z(\beta,\mu)}{Z(\beta,0)}
= \frac{1}{Z(\beta,0)}
\int \mathscr{D}U e^{i\theta(\mu)} \abs{\det M(\mu)}^{\Nf} e^{6\beta\Nsite P}
= \int dPdF \ave{e^{i\theta(\mu)}}_{(P,F)}
w_0(P,F,\beta,\mu),
\end{equation}
where the probability distribution function,
\begin{equation}
w_0(P',F',\beta,\mu) = \frac{1}{Z(\beta,0)}
\int \mathcal{D}U \delta(P'-P[U])\delta(F'-F[U])
\abs{\det M(\mu)}^{N_f}e^{6\beta \Nsite P},
\end{equation}
is obtained by the histogram of $P$ and $F$
in the phase quenched simulation, and
\begin{eqnarray}\label{eq:phase}
\ave{e^{i\theta(\mu)}}_{(P',F')}
&=& \frac{
\int\mathscr{D}U {e^{i\theta(\mu)}\delta(P'-P[U])\delta(F'-F[U])}
\abs{\det M(\mu)}^{N_f}e^{6\beta \Nsite P}}
{\int\mathscr{D}U \delta(P'-P[U])\delta(F'-F[U])
\abs{\det M(\mu)}^{N_f}e^{6\beta \Nsite P}} \nonumber \\
&=& \frac{
\Ave{e^{i\theta(\mu)}\delta(P'-P[U])\delta(F'-F[U])}_{(\beta,\mu)}}
{\Ave{\delta(P'-P[U])\delta(F'-F[U])}_{(\beta,\mu)}}
\end{eqnarray}
is the expectation value of the complex phase of the quark determinant
with fixed $P'$ and $F'$.
The double bracket indicates the expectation value in the phase quenched simulation.
Here, $\Nsite=N_s^3 \times N_t$ is the number of lattice sites and $\beta=6/g^2$.
Note that $\ave{e^{i\theta}}$ does not depend on $\beta$ since we can
factor out $e^{6\beta \Nsite P}$ from both the numerator and the denominator in \Eq{eq:phase}.
Introducing the effective potential $V_0 = - \ln w_0$,
the ratio of the partition function, \Eq{eq:part_func}, can be written as
\begin{equation}\label{eq:int_eff_pot}
\frac{Z(\beta,\mu)}{Z(\beta,0)}
= \int dPdF e^{-\ClB{V_0(P,F,\beta,\mu)
               - \ln \langle e^{i\theta(\mu)} \rangle_{(P,F)}}} \\
= \int dPdF e^{-V(P,F,\beta,\mu)}, \quad
V = V_0 - \ln \langle e^{i\theta(\mu)} \rangle.
\end{equation}

\begin{wrapfigure}[12]{r}{0.38\textwidth}
\resizedgraphics{0.38}{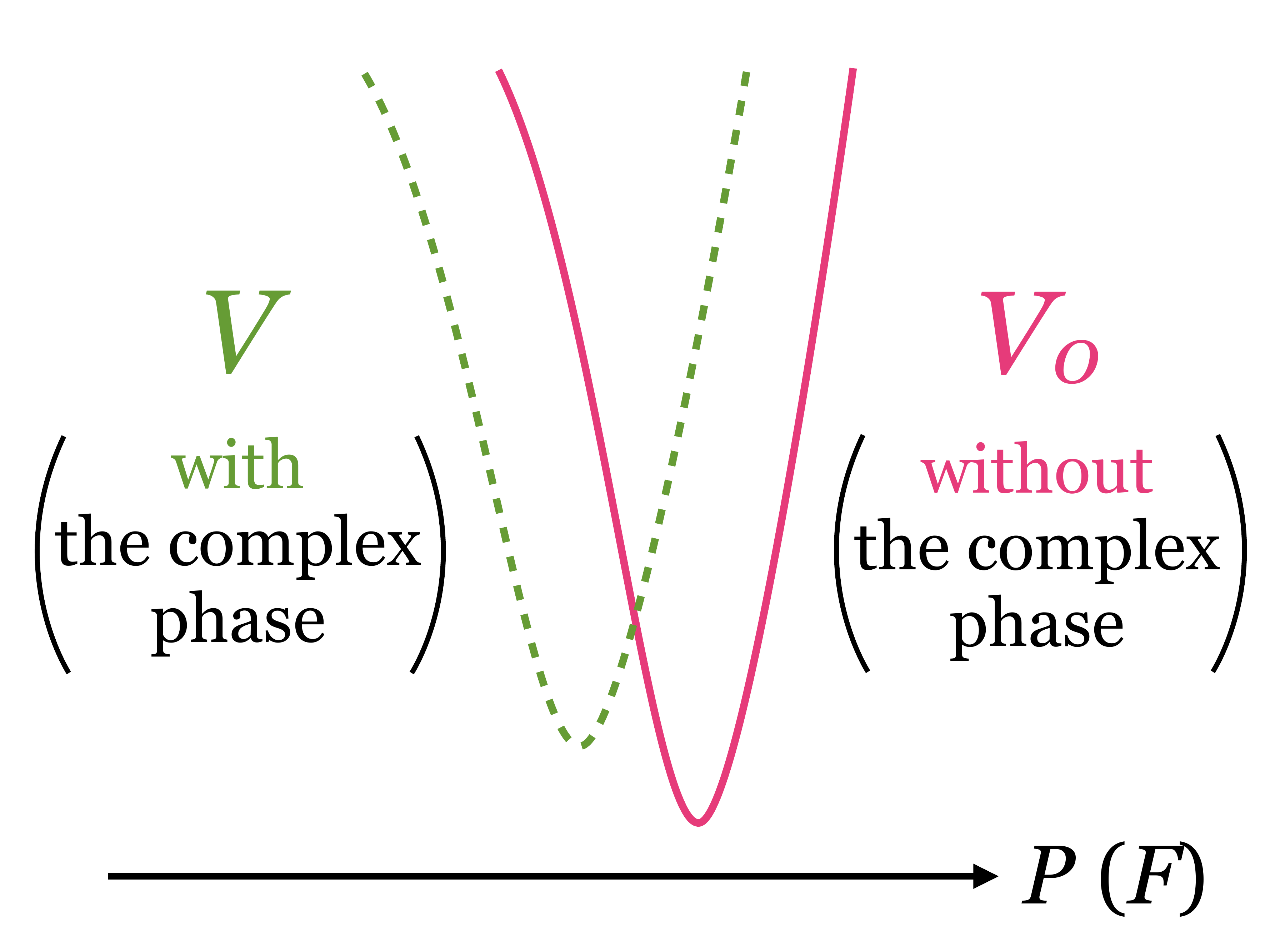}
\vspace{-5mm}
\caption{
A schematic figure for the overlap problem.
}
\label{fig:overlap_problem}
\end{wrapfigure}
In order to calculate the partition function precisely,
we have to evaluate the $P$ and $F$ integration in \Eq{eq:int_eff_pot} accurately:
namely, we have to generate gauge configurations near the minimum of $V$,
which dominates the integral in \Eq{eq:int_eff_pot}.
The histogram of $P$ and $F$ at a single simulation point $(\beta,\mu)$
provides the effective potential $V_0$ covering a limited region
in the $(P,F)$ plane.
If the complex phase (\ref{eq:phase}) has a large $P$ and/or $F$ dependence,
the minimum of $V$ differs from that of $V_0$, and we do not have
sufficient number of configurations near the minimum of $V$
(see \Fig{fig:overlap_problem}).
In such a case, there is no sufficient overlap between the important region for
the integral of \Eq{eq:int_eff_pot} and the region where $V_0$ is evaluated precisely
by measuring the histogram in the phase quenched simulation.

This overlap problem can be circumvented by combining the histograms
at several simulation points with the aid of the reweighting method.
The effective potential $V_0$ at $(\beta,\mu)$ can be obtained
by that at $(\beta_0,\mu_0)$ as follows:
\begin{equation}\label{eq:eff_pot_reweighting}
V_0(P,F,\beta,\mu) = V_0(P,F,\beta_0,\mu_0) -\ln R(P,F,\beta,\beta_0,\mu,\mu_0).
\end{equation}
Here $R(P',F',\beta,\beta_0,\mu,\mu_0) = w_0(P,F,\beta,\mu)/w_0(P,F,\beta_0,\mu_0)$
is the reweighting factor,
\begin{equation}\label{eq:reweighting}
R(P',F',\beta,\beta_0,\mu,\mu_0)
= e^{6(\beta-\beta_0)\Nsite P'}\frac{
\Ave{\delta(P'-P[U])\delta(F'-F[U])
\abs{\frac{\det M(\mu)}{\det M(\mu_0)}}^{\Nf}}_{(\beta_0,\mu_0)}
}{
\Ave{\delta(P'-P[U])\delta(F'-F[U])}_{(\beta_0,\mu_0)}}.
\end{equation}
Determination of $R$ is reduced to that of the expectation value of
the quark determinant in the phase quenched simulation when $\beta=\beta_0$.
Under a $\beta$ shift with keeping $\mu=\mu_0$,
the slope of the effective potential changes by a constant factor
while the curvature remains the same.
Combining the effective potentials $V_0$ at various simulation points
using \Eq{eq:eff_pot_reweighting},
we can have an enough overlap with the minimum of $V$
even if the complex phase has a large $P$ and/or $F$ dependence.
We can thus avoid the overlap problem.

Because the phase quenched simulations in two-flavor QCD correspond to
the case of isotriplet chemical potentials, a comment is in order about
the influence of the pion condensed phase.
The large isotriplet chemical potential induces the pion condensation
\cite{Son:2000xc}.
In the pion condensed phase $\ave{e^{i\theta}}_{(P,F)}$ is expected to vanish
as has been suggested in model calculations
\cite{Han:2008xj,Sakai:2010kx}.
This implies that $V_0(P,F)$ and $V(P,F) = V_0(P,F) - \ln\ave{e^{-i\theta}}_{(P,F)}$
have no overlap inside the condensed phase
and the partition function $Z(\beta,\mu)$ is dominated by configurations
outside the condensed phase.
Thus, we do not need to generate configurations with $\ave{e^{i\theta}}_{(P,F)}=0$,
which have no contribution to the integral in \Eq{eq:part_func}.

\vspace{-2mm}
\section{Cumulant expansion for the complex phase of the quark determinant}
\vspace{-1mm}

Even though we can circumvent the overlap problem,
a large fluctuation of the phase of the quark determinant at large chemical potential
leads to a frequent change of the sign of the complex phase.
In this case, Monte Carlo simulations suffer from the sign problem.
We exploit the cumulant expansion for the expectation value
of the complex phase to avoid the sign problem,
\begin{equation}
\langle e^{i\theta(\mu)} \rangle_{(P,F)}
= \exp\left[ i\ave{\theta}_c
- \frac{1}{2}\ave{\theta^2}_c
- \frac{i}{3!}\ave{\theta^3}_c
+ \frac{1}{4!}\ave{\theta^4}_c
+ \cdots \right].
\end{equation}
The cumulants agree with the central moments up to the third order,
while they differ at higher orders.
For instance, the fourth order cumulant is given by
\begin{equation}
\setlength{\abovedisplayskip}{6pt}
\setlength{\belowdisplayskip}{6pt}
\ave{\theta^4}_c
= \ave{\left(\theta-\ave{\theta}_{(P,F)}\right)^4}_{(P,F)}
- 3\ave{\left(\theta-\ave{\theta}_{(P,F)}\right)^2}^2_{(P,F)}.
\end{equation}
The odd-order cumulants change the sign under the flip of the sign of the chemical potential,
$\mu \leftrightarrow -\mu$, which transforms quarks into antiquarks.
Accordingly, only the even-order cumulants survive if the system is invariant
under this transformation, and the complex phase now becomes
\begin{equation}
\langle e^{i\theta(\mu)} \rangle_{(P,F)}
= \exp\left[ - \frac{1}{2}\ave{\theta^2}_c
+ \frac{1}{4!}\ave{\theta^4}_c + \cdots \right].
\end{equation}
We stress that the right-hand side is real and positive
once we drop the odd-order cumulants from the symmetry.
By applying the cumulant expansion for the complex phase,
the sign problem is reduced to the convergence problem of
the cumulant expansion; namely,
we have no sign problem if the cumulant expansion converges.

An ideal case for the convergence of the cumulant expansion
is the case that the phase $\theta$ has a Gaussian distribution.
In such a case, only the second-order cumulant survives,
$\langle e^{i\theta(\mu)} \rangle_{(P,F)}
= \exp\left[ - \ave{\theta^2}_{(P,F)}/2 \right]$.
If we calculate $\theta(\mu) = \Nf\Im[\ln\det M(\mu)]$
in the limited range $[-\pi,\pi)$ taking into account the periodicity of
the complex phase $\ave{e^{i\theta}}$,
the phase distribution may have no resemblance to the Gaussian distribution.
It is essential for the convergence of the cumulant expansion
to calculate the phase of the quark determinant
such that the distribution takes of nearly a Gaussian form.
In this study, we define the phase in the range $-\infty < \theta < \infty$.
Instead of calculating $\det M(\mu)$ directly,
we measure the the derivatives of $\ln\det M(\mu)$ with respect to $\mu$,
and then calculate the phase of the quark determinant by integrating
the derivatives over $\mu$,
\begin{equation}\label{eq:mu_integral_theta}
\theta(\mu) = \Nf \Im \left[ \ln \det M(\mu) \right]
            = \Nf \int^{\mu/T}_0 \Im \left[
              \frac{\partial(\ln\det M(\mu))}{\partial(\mu/T)}
              \right]_{\bar{\mu}} d\left( \frac{\bar{\mu}}{T} \right).
\end{equation}
Note that this is not a Taylor expansion; namely,
it is applicable to any values of the chemical potential.
Conventional phase in the range $[-\pi,\pi)$ is recovered
by taking the principal value of $\theta$ with the period of $2\pi$.
The integrand in \Eq{eq:mu_integral_theta},
\begin{equation}
\frac{\partial(\ln\det M(\mu))}{\partial(\mu/T)}
= \Tr \left( M^{-1}(\mu)\frac{\partial M(\mu)}{\partial(\mu/T)} \right),
\end{equation}
can be regarded as the sum of the local density operator defined at each lattice site.
If the density operator has a small correlation length
compared to the spatial size of the system,
we expect the Gaussian distribution for the operator
due to the central limit theorem \cite{Ejiri:2007ga}.
The volume dependence of the convergence of cumulant expansion
has been discussed in \cite{Ejiri:2009hq}.

The absolute value of the quark determinant, which is used to label
the gauge configurations, and the ratio of the quark determinant,
which is needed to evaluate the reweighting factor \Eq{eq:reweighting},
can be also obtained by integrating the real part
of the derivatives without further computational costs,
\begin{eqnarray}
           \label{eq:mu_integral_F}
F(\mu) &=& \Nf \ln \abs{\frac{\det M(\mu)}{\det M(0)}}
        =  \Nf \int^{\mu/T}_0 \Re \left[
           \frac{\partial(\ln\det M(\mu))}{\partial(\mu/T)}
           \right]_{\bar{\mu}} d\left( \frac{\bar{\mu}}{T} \right), \\
           \label{eq:mu_integral_C}
C(\mu) &=& \Nf \ln \abs{\frac{\det M(\mu)}{\det M(\mu_0)}}
        =  \Nf \int^{\mu/T}_{\mu_0} \Re \left[
           \frac{\partial(\ln\det M(\mu))}{\partial(\mu/T)}
           \right]_{\bar{\mu}} d\left( \frac{\bar{\mu}}{T} \right).
\end{eqnarray}
We note that $\theta(\mu)$, $F(\mu)$, and $C(\mu)$ can be obtained
as continuous functions of $\mu$ in this approach.
In addition, the statistical errors for the reweighting factor $R$ are expected
to be small for fixed $F$ since $F$ and $C$ are strongly correlated.

\vspace{-2mm}
\section{Numerical simulations and the results}
\vspace{-1mm}

In this study, we use the RG-improved Iwasaki action for gauge action
and the $\Nf = 2$ $O(a)$-improved Wilson quark action with
$c_{SW} = ( 1 - 0.8412 \beta^{-1} )^{-3/4}$.
The ratio of pseudoscalar and vector meson masses at $T=\mu=0$
are set to $m_{\textrm{PS}}/m_{\textrm{V}} = 0.8$.
We generate gauge configurations on $8^3\times 4$ lattice
with the complex phase of the quark determinant removed.
The measurement of the first and the second derivatives of $\ln \det M(\mu)$
with respect to $\mu$, which are used to interpolate the integrands
in \Eq{eq:mu_integral_theta}, (\ref{eq:mu_integral_F})
and (\ref{eq:mu_integral_C}), has been done every 10 trajectories.
We employ the random noise method of \cite{Ejiri:2009hq} with 50 noises.
The statistics is the order of $O(1000)$.

\begin{figure}[htbp]
\begin{center}
\resizedgraphics{0.45}{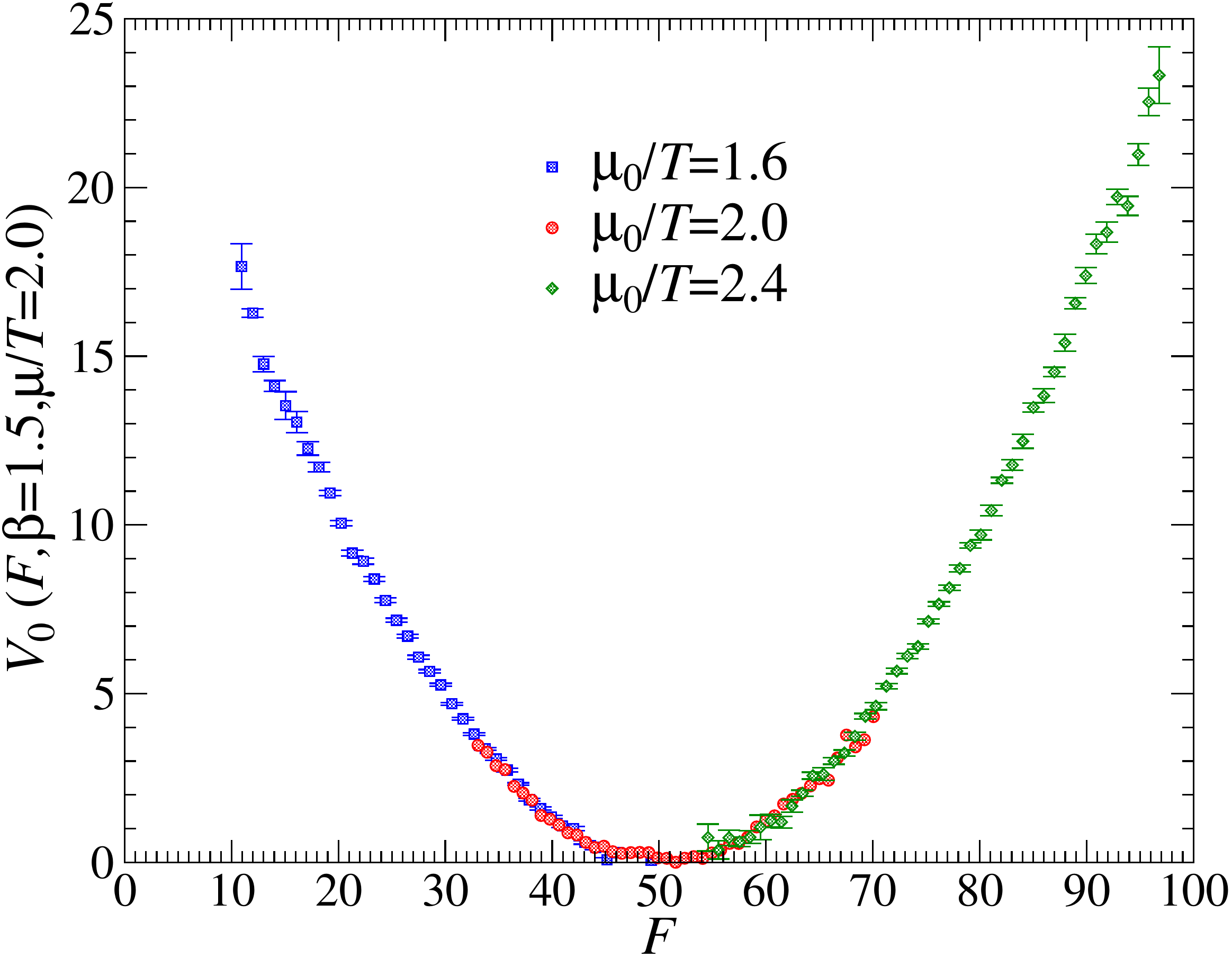}
\end{center}
\vspace{-5mm}
\caption{
The effective potential $V_0(F)$ at $\mu/T=2.0$
evaluated at three different simulation points.
}
\label{fig:prob_dist_func}
\end{figure}

The effective potential without the complex phase,
$V_0(F) = -\ln w_0(F)$, at $\mu/T=2.0$ evaluated at three different
simulation points, $(\beta,\mu_0/T)=(1.5,1.6)$, $(1.5,2.0)$, $(1.5,2.4)$,
is drawn in \Fig{fig:prob_dist_func}.
$V_0(F)$ is normalized such that $V_0(F)=0$ at the minimum
for each simulation.
We observe that the three data sets covering different ranges nicely fall on one curve.
Although $V_0$ at the single simulation point $(\beta,\mu_0/T) = (1.5, 2.0)$
covers only the narrow range centered around $F \sim 50$,
the effective potential obtained from the histograms at different simulation points
by the use of the reweighting method widely range in the $F$ direction.
Moreover, we see that the statistical errors of the effective potential,
which stem from the reweighting factor, are small as we expected.

\begin{figure}[htbp]
\begin{center}
\begin{minipage}{0.4\hsize}
\resizedgraphics{1.0}{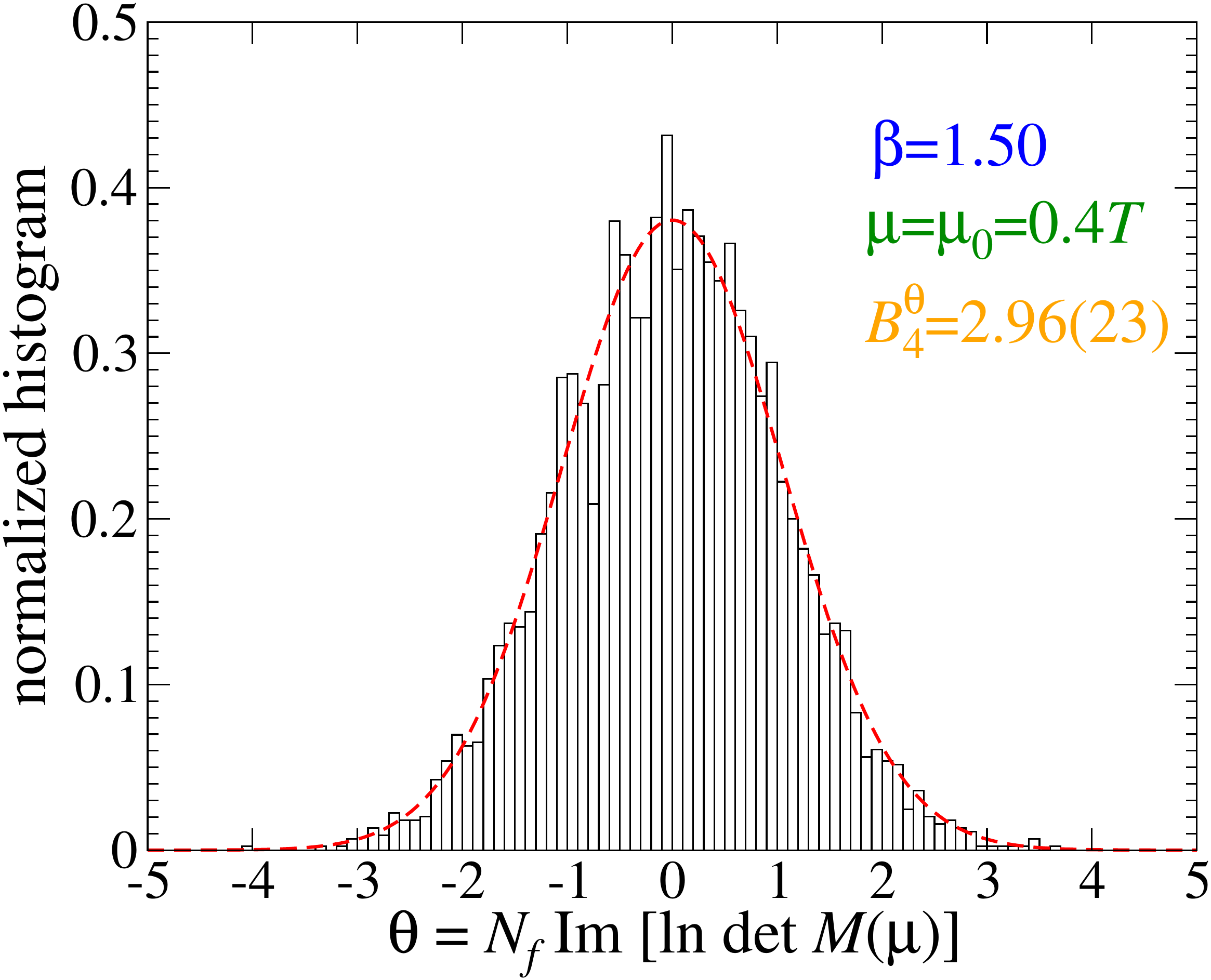}
\end{minipage}
\hspace{0.05\hsize}
\begin{minipage}{0.4\hsize}
\resizedgraphics{1.0}{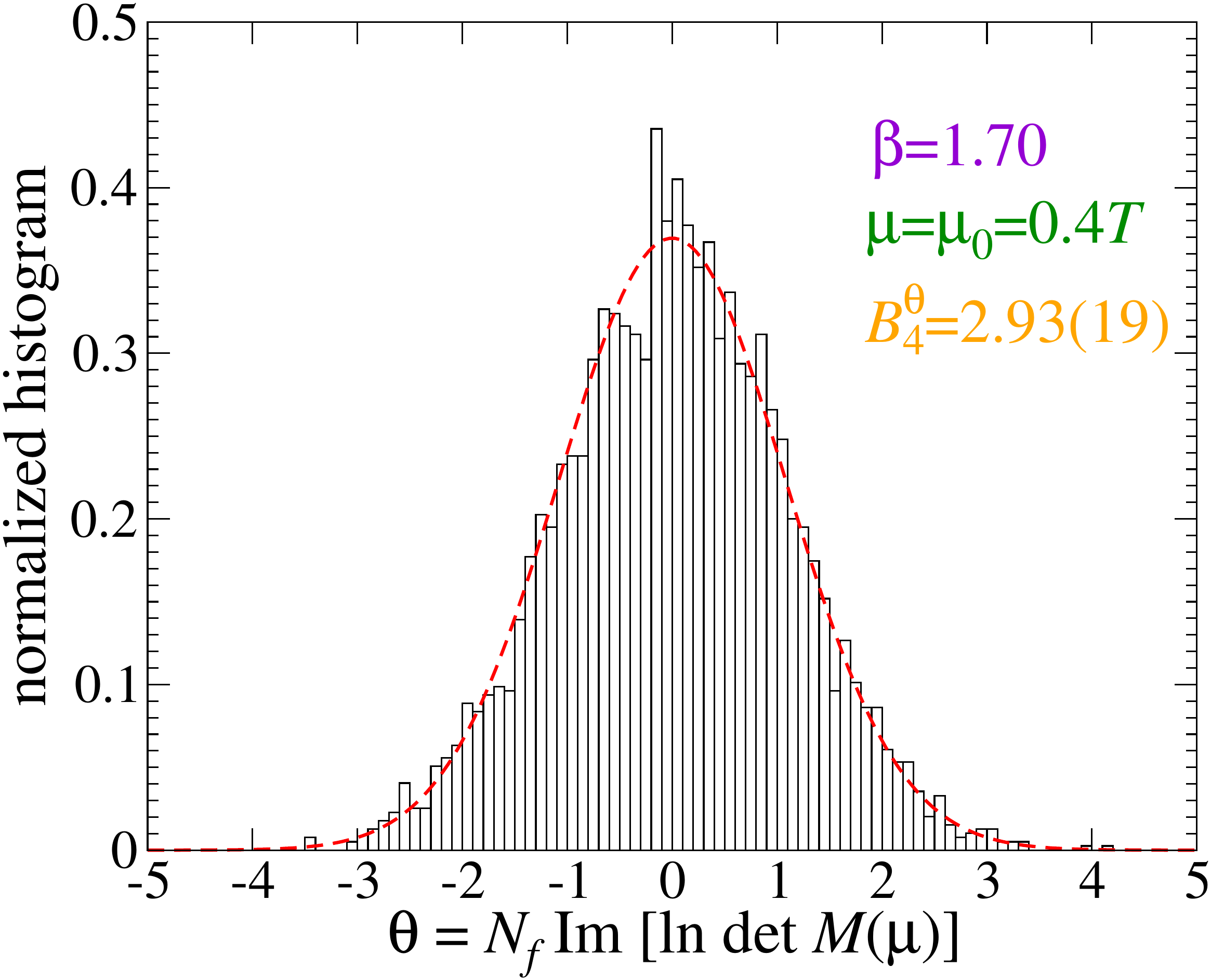}
\end{minipage}
\end{center}
\vspace{-10mm}
\begin{center}
\begin{minipage}{0.4\hsize}
\resizedgraphics{1.0}{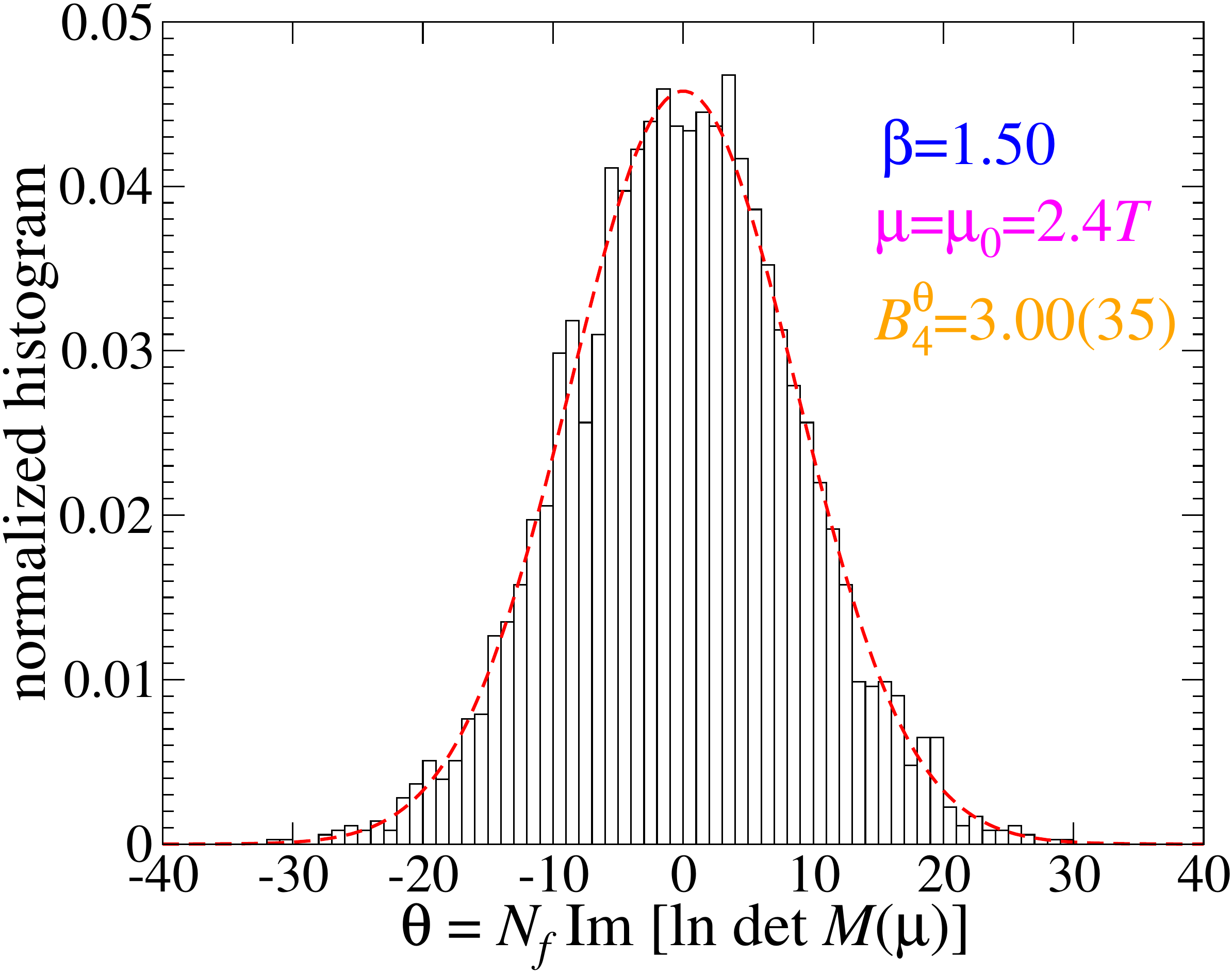}
\end{minipage}
\hspace{0.05\hsize}
\begin{minipage}{0.4\hsize}
\resizedgraphics{1.0}{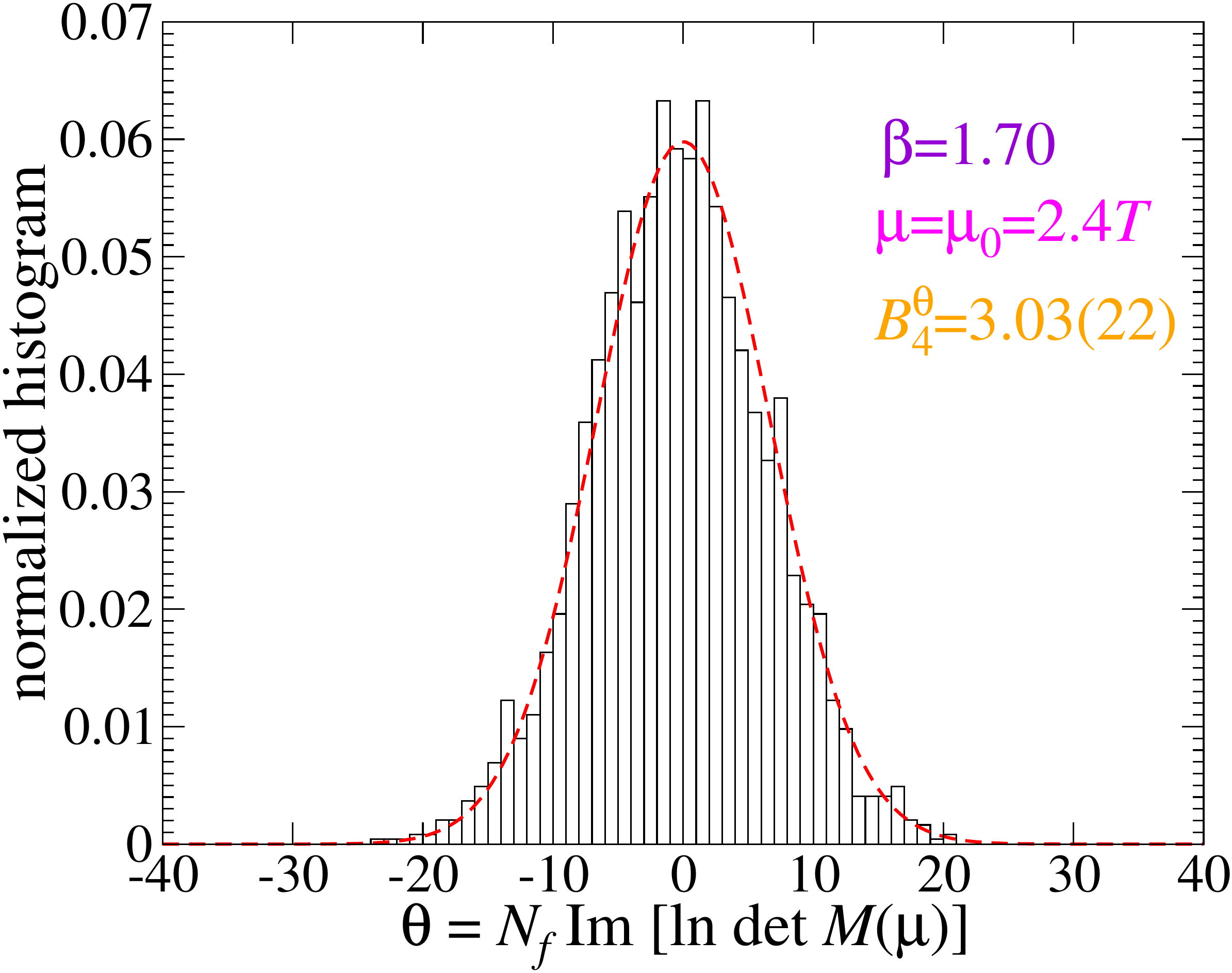}
\end{minipage}
\end{center}
\vspace{-8mm}
\caption{
The distribution of the phase of the quark determinant
at $\mu/T=0.4$ (upper panels) and $\mu/T=2.4$ (lower panels).
The dashed curves are the fitted results with the Gaussian function.
$B_4^{\theta}$ is the fourth-order Binder cumulant normalized
such that $B_4^{\theta}=3$ for the Gaussian function.
}
\label{fig:phase_distribution}
\end{figure}

The distribution of the phase of the quark determinant is
depicted in \Fig{fig:phase_distribution}.
The dashed curves are the fitted results with a Gaussian function.
$B_4^{\theta}$ is the fourth-order Binder cumulant normalized
such that $B_4^{\theta}=3$ for a Gaussian distribution, i.e.,
$B_4^{\theta} \equiv \langle\theta^4\rangle_c/\langle\theta^2\rangle_c^2 +3$.
The upper and lower panels are the results at $\mu/T = 0.4$ and $2.4$, respectively.
We see that the phase distribution gets broader as
the chemical potential increases.
Furthermore, at large chemical potential (lower panels),
we observe that the phase distributes broadly at low temperature.
The important point is that the phase distribution evaluated by
\Eq{eq:mu_integral_theta} can be well approximated
by a Gaussian function even in the high density region $\mu/T > 1$.
This promises a good convergence in the cumulant expansion
for the complex phase of the quark determinant.

\begin{figure}[htdp]
\begin{center}
\begin{minipage}{0.4\hsize}
\resizedgraphics{1.}{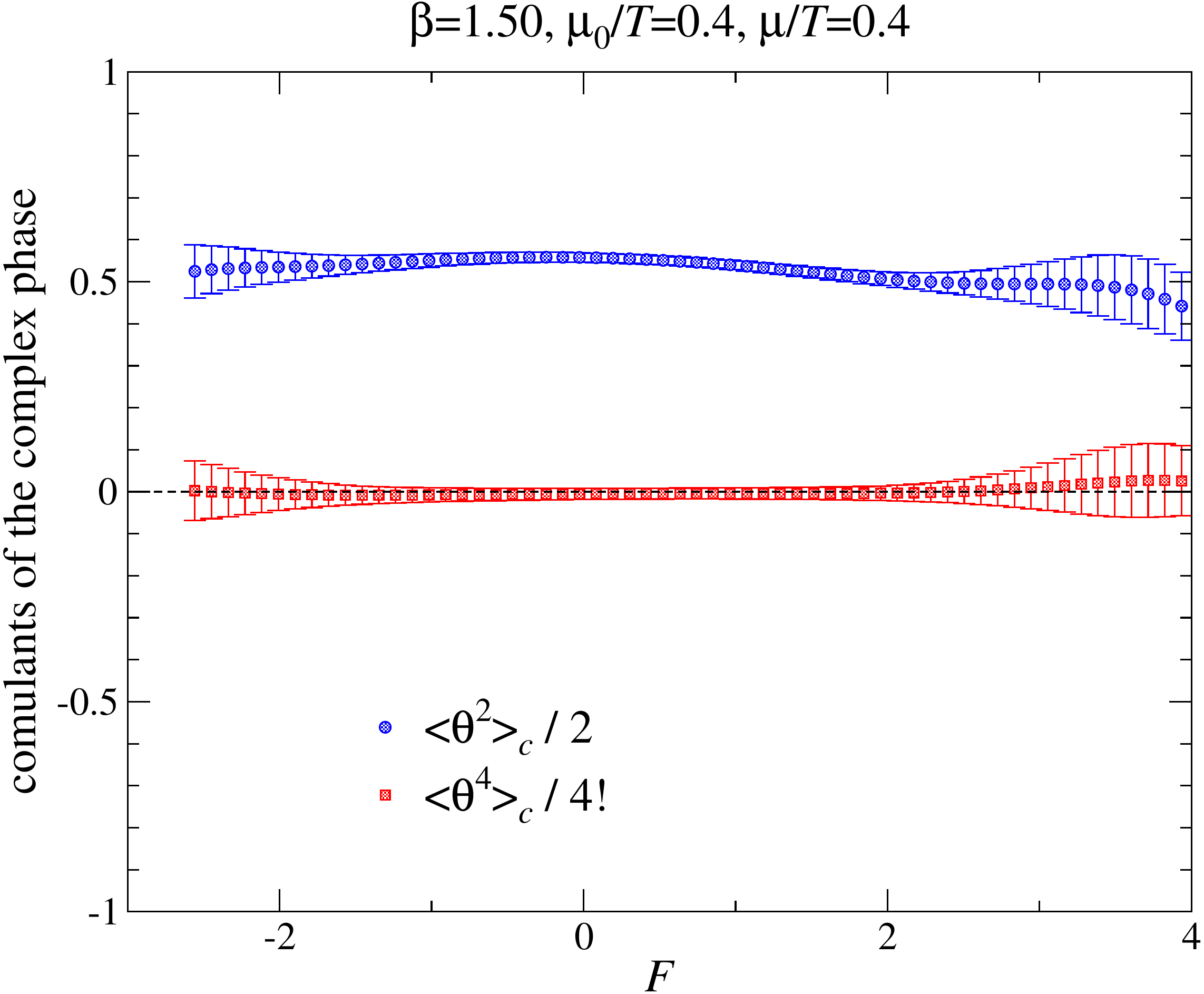}
\end{minipage}
\hspace{0.05\hsize}
\begin{minipage}{0.4\hsize}
\resizedgraphics{1.}{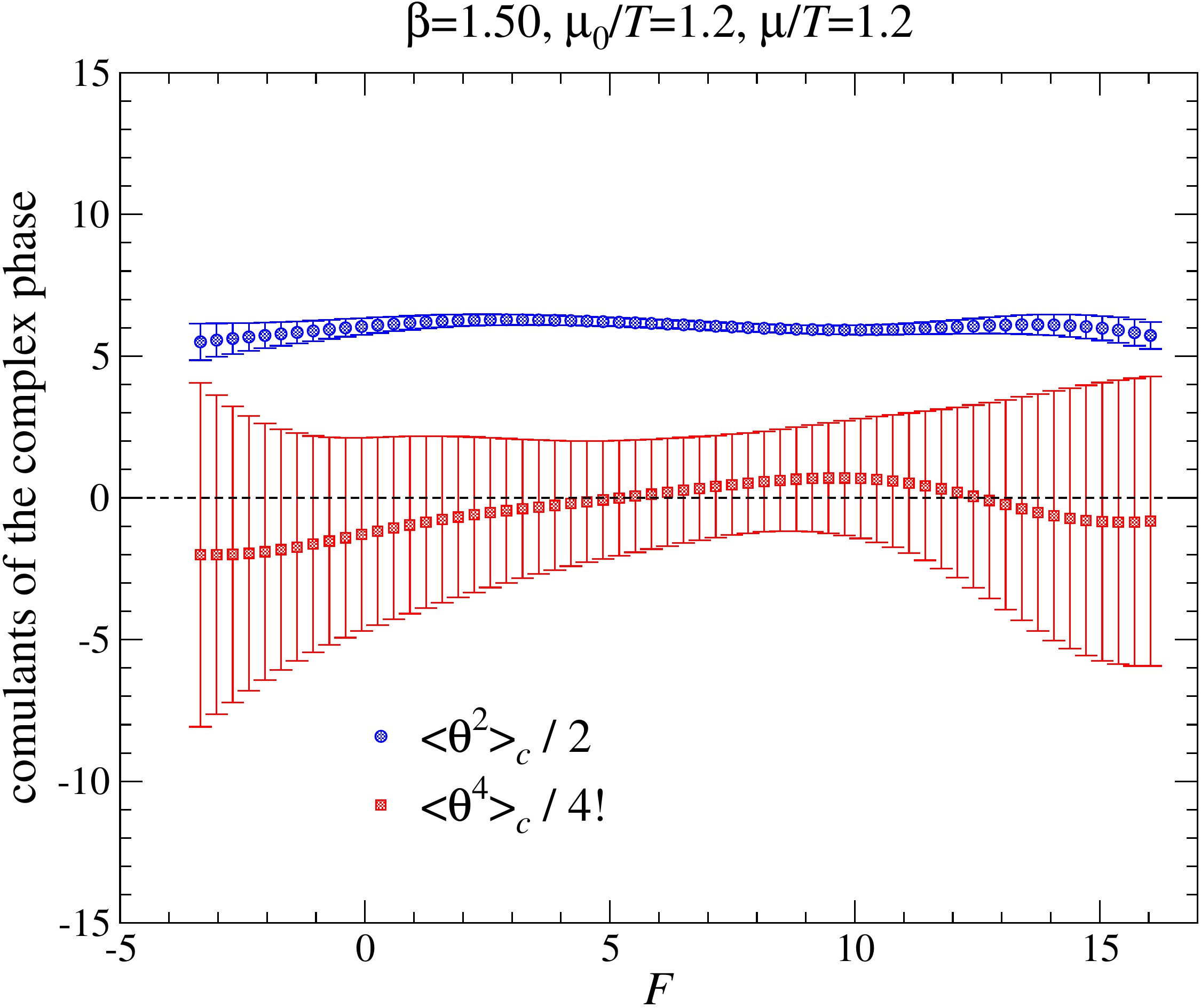}
\end{minipage}
\end{center}
\vspace{-10mm}
\begin{center}
\begin{minipage}{0.4\hsize}
\resizedgraphics{1.}{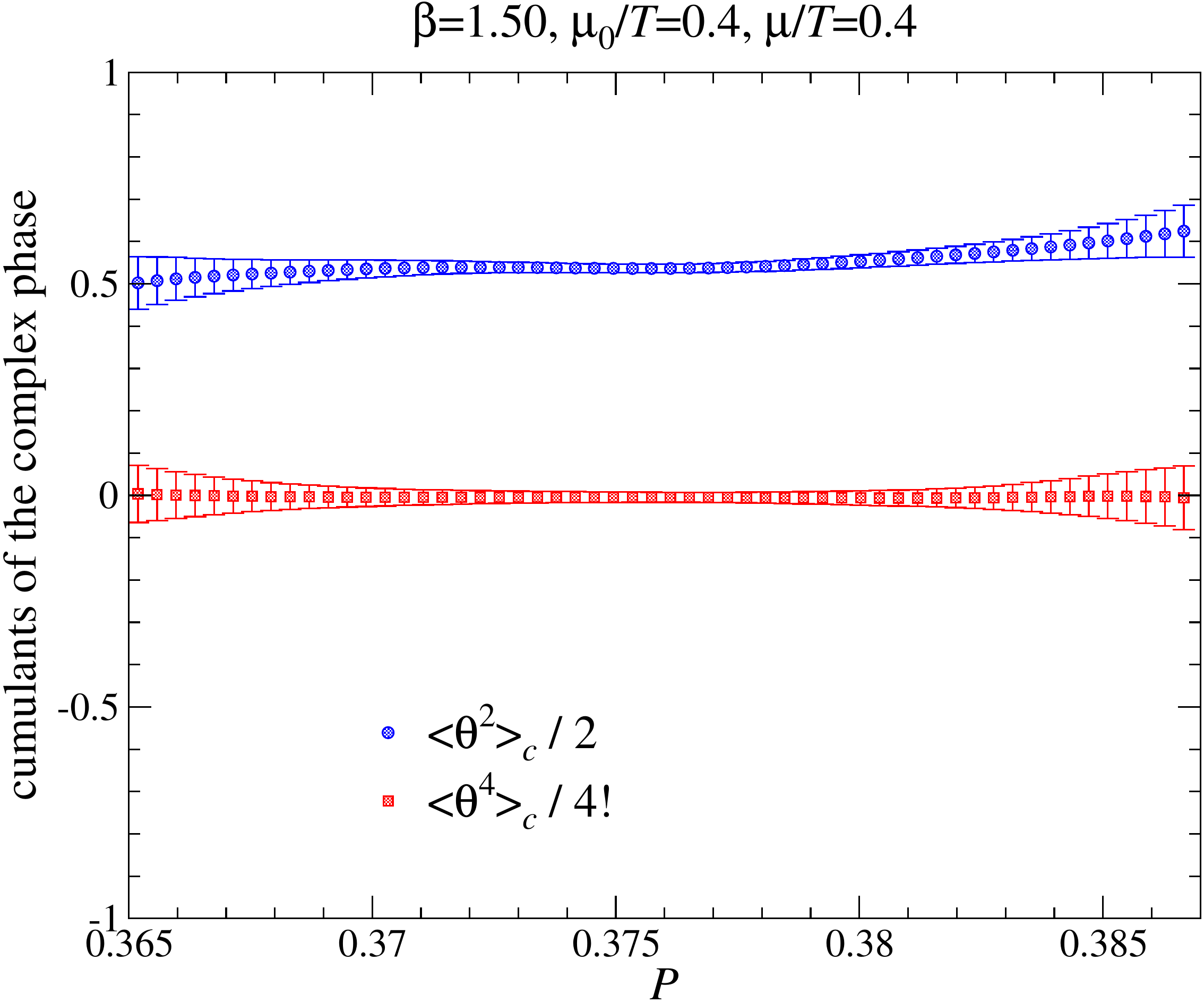}
\end{minipage}
\hspace{0.05\hsize}
\begin{minipage}{0.4\hsize}
\resizedgraphics{1.}{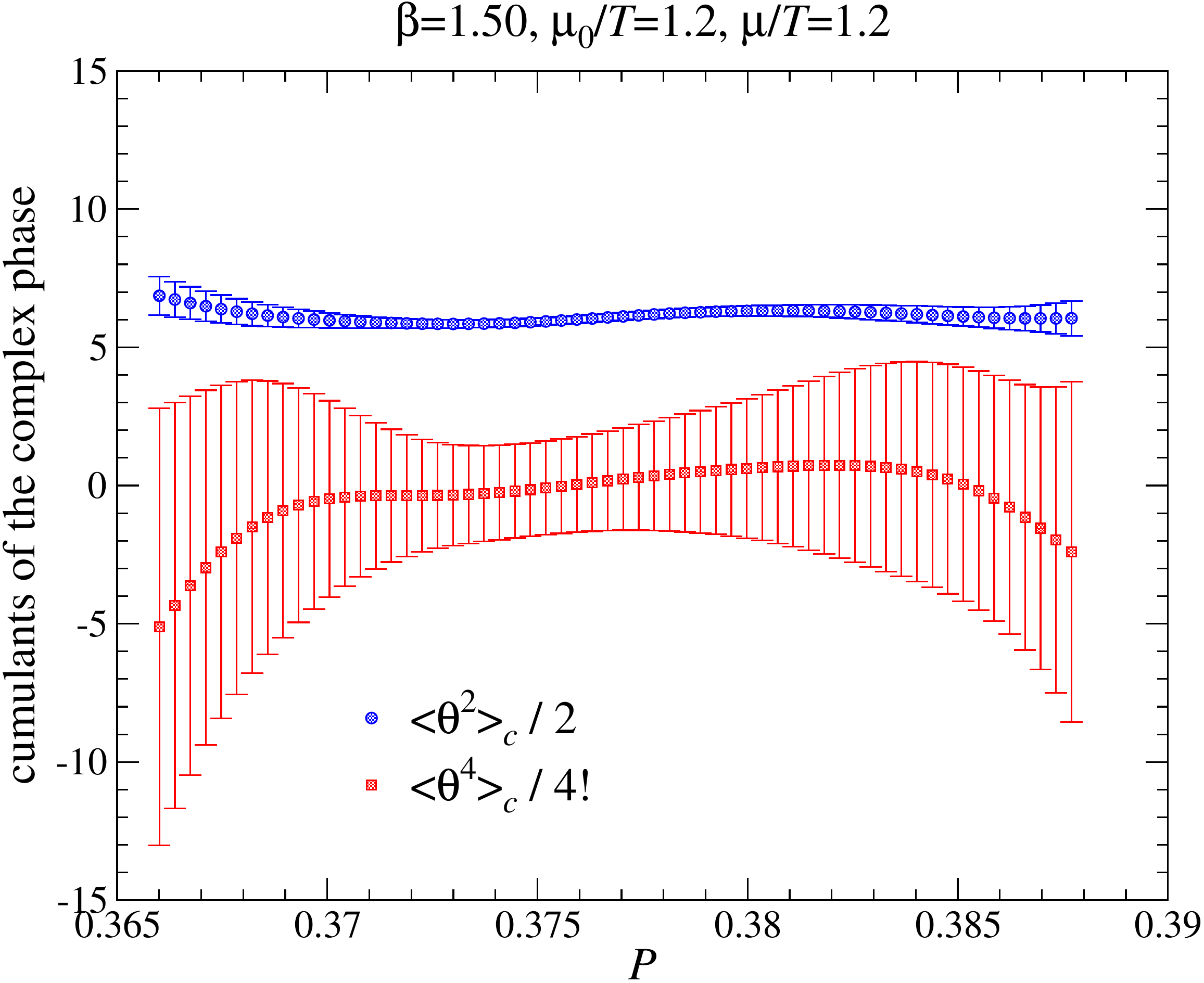}
\end{minipage}
\end{center}
\vspace{-8mm}
\caption{
The second and the fourth order of the cumulants
as a function of $F$ (upper panels) and $P$ (lower panels)
at $\mu/T=\mu_0/T=0.4$ (left panels) and $\mu/T=\mu_0/T=1.2$ (right panels).
}
\label{fig:cumulants}
\end{figure}

\Fig{fig:cumulants} shows the second and the fourth order of the cumulants
as a function of $F$ (upper panels) and $P$ (lower panels)
at $\mu/T=\mu_0/T=0.4$ (left panels) and $\mu/T=\mu_0/T=1.2$ (right panels).
We observe that the second order cumulant increases with $\mu/T$
both as a function of $F$ and $P$.
On the other hand, the fourth order cumulant is consistent with zero
within the statistical errors although the errors grow with $\mu/T$.
We do not see a clear $F$ and $P$ dependence of the second and fourth cumulants
in these parameter region.

\vspace{-2mm}
\section{Summary}
\vspace{-1mm}

In this study, we proposed a new approach to finite density QCD
based on the histogram method and the reweighting technique
with phase quenched simulations.
We apply the cumulant expansion for the complex phase of the quark determinant.
We found that the reweighting technique combined with the histogram method
works well and we obtained the effective potential covering a wide range.
Moreover, we proposed a way to calculate the phase $\theta$
of the quark determinant, which is not constrained in the range $[-\pi,\pi)$.
We showed that the distribution of $\theta$ becomes wide as $\mu/T$ increases
and the phase distribution can be well approximated
by a Gaussian function both at small and large chemical potential.
The second order cumulant increases with the chemical potential,
while the fourth order cumulant is consistent with zero
although the statistical error increases with $\mu$.
A comprehensive analysis in a wide parameter region is in progress.

\vspace{2mm}

This work is in part supported by Grants-in-Aid of the Japanese Ministry of Education,
Culture, Sports, Science and Technology,
(Nos.20340047, 21340049, 22740168, 23540295)
and by the Grant-in-Aid for Scientific Research on Innovative Areas
(Nos. 20105001, 20105003, 2310576).
HO is supported by the Japan Society for the Promotion of Science for
Young Scientists.


\end{document}